\documentclass[11pt]{article}

\usepackage[utf8x]{inputenc}

\usepackage{float}
\usepackage{graphicx}
\usepackage{caption}
\usepackage{subcaption}
\usepackage{geometry}
\usepackage{amsmath}
\usepackage{url}

\usepackage{amsfonts}
\usepackage{amssymb}
\usepackage{hyperref}
\usepackage{booktabs}
\usepackage{multirow}
\usepackage{makecell}
\usepackage{authblk}
\usepackage{bm}
\usepackage{enumitem}
\setlist[enumerate]{itemsep=0mm}
\usepackage[round,authoryear,comma,sort&compress,numbers]{natbib}

\usepackage{verbatim} 

\begin{document}

\title{Multiscale Information Decomposition: Exact Computation for Multivariate Gaussian Processes}
\author{L. Faes}

\affil{%
 Bruno Kessler Foundation, Trento, Italy
}%
\affil{
 BIOtech, Dept. of Industrial Engineering, University of Trento, Italy
}


\author{S. Stramaglia}
\affil{
 Dipartimento di Fisica, Universit\'a degli Studi Aldo Moro, Bari, Italy
}%
\affil{
 INFN, Sezione di Bari, Italy
}%
\author{D. Marinazzo}
\affil{%
 Data Analysis Department, Ghent University, Ghent, Belgium
}%

\fontfamily{ptm}\selectfont

\maketitle

\begin{abstract}Exploiting the theory of state space models, we derive the exact expressions of the information transfer, as well as redundant and synergistic transfer, for coupled Gaussian processes observed at multiple temporal scales. All the terms, constituting the frameworks known as {\it interaction information decomposition} and {\it partial information decomposition}, can thus be analytically obtained for different time scales from the parameters of the VAR model that fits the processes. 
We report the application of the proposed methodology firstly to benchmark Gaussian systems, showing that this class of systems  may generate patterns of information decomposition characterized by prevalently redundant or synergistic information transfer persisting across multiple time scales, or even by alternating prevalence of redundant and synergistic source interaction depending on the time scale. Then, we apply our method to an important topic in neuroscience, i.e. the detection of causal interactions in human epilepsy networks, for which we show the relevance of partial information decomposition to the detection of multiscale information transfer spreading from the seizure onset zone.
\end{abstract}

\section{Introduction}

The information-theoretic treatment of groups of correlated degrees of freedom can reveal their functional roles as memory structures or information processing units. A large body of recent work has shown how the general concept of ``information processing'' in a network of multiple interacting dynamical systems described by multivariate stochastic processes can be dissected into basic elements of computation defined within the so-called framework of information dynamics \cite{lizier2014framework}. These elements essentially reflect the new information produced at each moment in time about a target system in the network \cite{pincus1995approximate}, the information stored in the target system \cite{lizier2012local, wibral2014local}, the information transferred to it from the other connected systems \cite{schreiber2000measuring, wibral2014directed} and the modification of the information flowing from multiple source systems to the target \cite{lizier2010information, wibral2015bits}. The measures of information dynamics have gained more and more importance in both theoretical and applicative studies in several fields of science \cite{lizier2011information, wibral2011transfer, hlinka2013reliability, barnett2013information, marinazzo2014information, faes2014information, faes2015information, porta2015conditional, faes2017information, wollstadt2017breakdown}. While the information-theoretic approaches to the definition and quantification of new information, information storage and information transfer are well understood and widely accepted, the problem of defining, interpreting and using measures of information modification has not been fully addressed in the~literature.

Information modification in a network is tightly related to the concepts of redundancy and synergy between source systems sharing information about a target system, which refer to the existence of common information about the target that can be retrieved when the sources are used separately (redundancy) or when they are used jointly (synergy) \cite{schneidman2003synergy}. Classical multivariate entropy-based approaches refer to the {interaction information decomposition} (IID), which reflects information modification through the balance between redundant and synergetic interaction among different source systems influencing the target \cite{stramaglia2012expanding, stramaglia2014synergy,StramagliaIEEE}.
The IID framework has the drawback that it implicitly considers redundancy and synergy as mutually exclusive concepts, because it quantifies information modification with a single measure of interaction information \cite{mcgill1954multivariate} (also called co-information \cite{bell2003co}) that takes positive or negative values depending on whether the net interaction between the sources is synergistic or redundant.
This limitation has been overcome by the elegant mathematical framework introduced by Williams and Beer \cite{williams2010nonnegative}, who proposed the so-called {partial information decomposition} (PID) as a nonnegative decomposition of the information shared between a target and a set of sources into terms quantifying separately unique, redundant and synergistic contributions. However, the PID framework has the drawback that the terms composing the PID cannot be obtained unequivocally from classic measures of information theory (i.e., entropy and mutual information), but a new definition of either redundant, synergistic or unique information needs to be provided to implement the decomposition. 
Accordingly, much effort has {focused on} finding the most proper measures to define the components of the PID, with alternative proposals defining new measures of redundancy \cite{williams2010nonnegative,harder2013bivariate}, synergy \cite{griffith2014intersection, e19020085} or unique information \cite{bertschinger2014quantifying}. The proliferation of~different definitions is mainly due to the fact that there is no full consensus on which axioms should be stated to impose desirable properties for the PID measures. An additional problem which so far has seriously limited the practical implementation of these concepts is the difficulty in providing reliable estimates of the {information measures appearing in the IID and PID decompositions.} The naive estimation of probabilities by histogram-based methods followed by the use of plug-in estimators leads to serious bias problems \cite{panzeri2007correcting, faes2014conditional}. While the use of binless density estimators \cite{kozachenko1987sample} and the adoption of schemes for dimensionality reduction \cite{vlachos2010nonuniform, marinazzo2012causal} have been shown to improve the reliability of estimates of~information storage and transfer \cite{Faes2015estimatingthe}, the effectiveness of these approaches for the computation of~measures of information modification has not been demonstrated yet. Interestingly, both the problems of defining appropriate PID measures and of reliably estimating these measures from data are much alleviated if one assumes that the observed variables have a joint Gaussian distribution. Indeed, in such a case, recent studies have proven the equivalence between most of the proposed redundancy measures to be used in the PID \cite{barrett2015exploration} and have provided closed form solutions to the issue of computing any measure of information dynamics from the parameters of the vector autoregressive (VAR) model that characterizes an observed multivariate Gaussian process \cite{barrett2010multivariate, porta2017quantifying, faes2017information}.

The second fundamental question that is addressed in this study is relevant to the computation of~information dynamics for stochastic processes displaying multiscale dynamical structures. It is indeed well known that many complex physical and biological systems exhibit peculiar oscillatory activities, which are deployed across multiple temporal scales \cite{Ivanov1999461,chou2011wavelet,wang2013multiscale}. The most common way to investigate such activities is to resample at different scales, typically through low pass filtering and downsampling \cite{costa2002multiscale, Valencia20092202}, the originally measured realization of an observed process, so as to yield a~set of rescaled time series, which are then analyzed employing different dynamical measures. This~approach is well established and widely used for the multiscale entropy analysis of individual time series measured from scalar stochastic processes. However, its extension to the investigation of the multiscale structure of the information transfer among coupled processes is complicated by {theoretical and practical} issues \cite{barnett2015granger, solo2016state}. Theoretically, the procedure of rescaling alters the causal interactions between lagged components of the processes in a way that is not fully understood and, if~not properly performed, may alter the temporal relations between processes and thus induce spurious detection of~information transfer. In practical analysis, filtering and downsampling are known to degrade severely the estimation of information dynamics and to impact consistently the detectability, accuracy and data demand \cite{florin2010effect, barnett2017detectability}. 

In recent works, we have started tackling the above problems within the framework of linear VAR modeling of multivariate Gaussian processes, with the focus on the multiscale computation of information storage and information transfer \cite{faes2016multiscale,faes2017multiscalegranger}. In this study, we aim at extending these recent theoretical advances to the multiscale analysis of information modification in multivariate Gaussian systems performed through the IID and PID decomposition frameworks.
To this end, we exploit the theory of~state space (SS) models \cite{Aoki1991} and build on recent theoretical results \cite{barnett2015granger, solo2016state} to show that exact values of~interaction transfer, as well as redundant and synergistic transfer can be obtained for coupled Gaussian processes observed at different time scales starting from the parameters of the VAR model that fits the processes and from the scale factor. The theoretical derivations are first used in examples of~benchmark Gaussian systems, reporting that these systems may generate patterns of information decomposition characterized by prevalently redundant or synergistic information transfer persisting across multiple time scales or even by alternating the prevalence of redundant and synergistic source interaction depending on the time scale. The high computational reliability of the SS approach is then exploited in the analysis of real data by the application to a topic of great interest in neuroscience, i.e., the detection of information transfer in epilepsy networks.

The proposed framework is implemented in the {msID} MATLAB\textsuperscript{\textregistered} toolbox, which is uploaded as Supplementary Material to this article and is freely available for download from \url{www.lucafaes.net/msID.html} and \url{https://github.com/danielemarinazzo/multiscale_PID}

\section{Information Transfer Decomposition in Multivariate Processes}\label{sec2}

Let us consider a discrete-time, stationary vector stochastic process composed of {\textit{M}} real-valued zero-mean scalar processes, $\bm{Y}_n=[Y_{1,n}\cdots Y_{M,n}]^T$, $-\infty < n < \infty$. In an information-theoretic framework, the information transfer between scalar sub-processes is quantified by the well-known transfer entropy (TE), which is a popular measure of the ``information transfer'' directed towards an~assigned {{target}} process from one or more {source} processes. Specifically, the TE quantifies the amount of~information that the past of the source provides about the present of the target over and above the information already provided by the past of the target itself \cite{schreiber2000measuring}. Taking $Y_j$ as target and $Y_i$ as source, the TE is defined as: 
\begin{equation} \label{eq:TE}
		\mathcal{T}_{i\rightarrow j} = I(Y_{j,n};Y_{i,n}^-|Y_{j,n}^-)
\end{equation}
where $Y_{i,n}^-=[Y_{i,n-1}Y_{i,n-2}\cdots]$ and $Y_{j,n}^-=[Y_{j,n-1}Y_{j,n-2}\cdots]$ represent the past of the source and target processes and $I(\cdot;\cdot|\cdot)$ denotes conditional mutual information (MI). In the presence of two sources $Y_i$ and $Y_k$ and a target $Y_j$, the information transferred toward $Y_j$ from the sources $Y_i$ and $Y_k$ taken together is quantified by the joint TE: 
\begin{equation} \label{eq:JTE}
		\mathcal{T}_{ik\rightarrow j} = I(Y_{j,n};Y_{i,n}^-,Y_{k,n}^-|Y_{j,n}^-) .
\end{equation}

Under the premise that the information jointly transferred to the target by the two sources is different than the sum of the amounts of information transferred individually, in the following, we present two possible strategies to decompose the joint TE into amounts eliciting the individual TEs, as well as redundant and/or synergistic TE terms.

\subsection{Interaction Information Decomposition}

The first strategy, which we denote as interaction information decomposition (IID), decomposes the joint TE (\ref{eq:JTE}) as:
\begin{equation} \label{eq:IID}
		\mathcal{T}_{ik\rightarrow j} = \mathcal{T}_{i\rightarrow j} + \mathcal{T}_{k\rightarrow j} + \mathcal{I}_{ik\rightarrow j},
\end{equation}
where $I_{ik\rightarrow j}$ is denoted as {interaction transfer entropy} (ITE) because it is equivalent to the interaction information \cite{mcgill1954multivariate} computed between the present of the target and the past of the two sources, conditioned to the past of the target:
\begin{equation} \label{eq:IIdef}
		\mathcal{I}_{ik\rightarrow j} = I(Y_{j,n};Y_{i,n}^-;Y_{k,n}^-|Y_{j,n}^-) .
\end{equation}

\begin{figure}[H]
\centering
\includegraphics[width=14 cm]{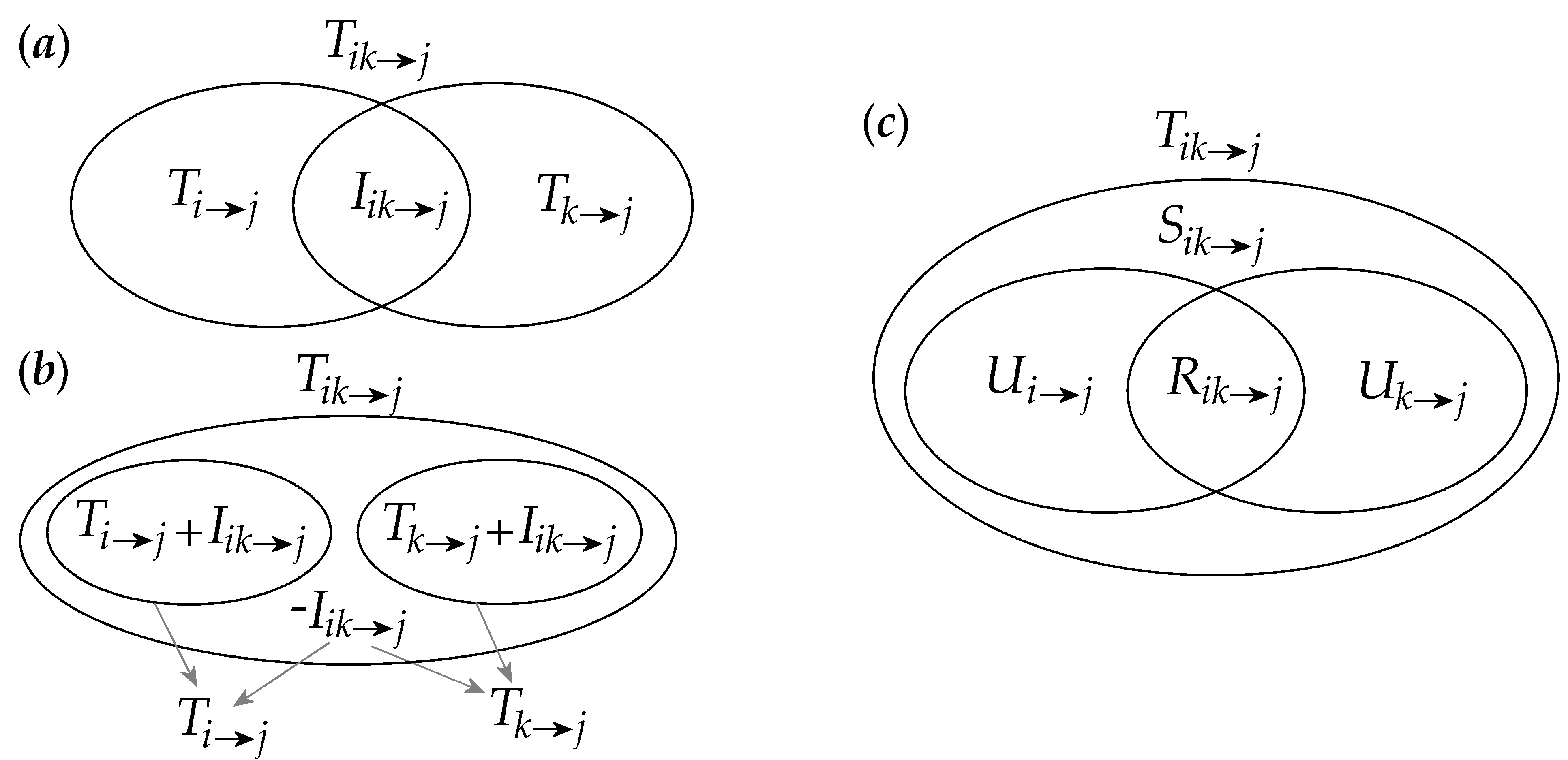}
\caption{Venn diagram representations of the interaction information decomposition (IID) (\textbf{{a}},\textbf{b})) and the partial information decomposition (PID) (\textbf{c})). The IID is depicted in a way such that all areas in the diagrams are positive: the interaction information transfer $\mathcal{I}_{ik\rightarrow j}$ is positive in ({a}), denoting net synergy, and is {negative in} ({b}), denoting net redundancy.}
\label{fig1}
\end{figure}

The interaction TE quantifies the modification of the information transferred from the source processes $Y_i$ and $Y_k$ to the target $Y_j$, being positive when $Y_i$ and $Y_k$ cooperate in a synergistic way and negative when they act redundantly.
This interpretation is evident from the diagrams of~Figure~\ref{fig1}: in the case of synergy (Figure~\ref{fig1}a), the two sources $Y_i$ and $Y_k$ taken together contribute to the target $Y_j$ with more information than the sum of their individual contributions $(\mathcal{T}_{ik\rightarrow j}>\mathcal{T}_{i\rightarrow j}+\mathcal{T}_{k\rightarrow j})$, and the ITE is positive; in the case of redundancy (Figure~\ref{fig1}b), the sum of the information amounts transferred individually from each source to the target is higher than the joint information transfer $(\mathcal{T}_{i\rightarrow j}+\mathcal{T}_{k\rightarrow j}>\mathcal{T}_{ik\rightarrow j})$, so that the ITE is negative.

\subsection{Partial Information Decomposition}

An alternative expansion of the joint TE is that provided by the so-called partial information decomposition (PID) \cite{williams2010nonnegative}. The PID evidences four distinct quantities measuring the unique information transferred from each individual source to the target, measured by the {unique TEs} 
$\mathcal{U}_{i\rightarrow j}$ and $\mathcal{U}_{k\rightarrow j}$, \mbox{and the redundant} and synergistic information transferred from the two sources to the target, measured by the {redundant TE}
$\mathcal{R}_{ik\rightarrow j}$ and the {synergistic TE} $\mathcal{S}_{ik\rightarrow j}$. These four measures are related to each other and to the joint and individual TEs by the {following equations} (see also Figure \ref{fig1}c): 

\begin{subequations} \label{eq:PID}
\begin{align}
		\mathcal{T}_{ik\rightarrow j} = \mathcal{U}_{i\rightarrow j} + \mathcal{U}_{k\rightarrow j} + \mathcal{R}_{ik\rightarrow j} + \mathcal{S}_{ik\rightarrow j},\\
		\mathcal{T}_{i\rightarrow j} = \mathcal{U}_{i\rightarrow j} + \mathcal{R}_{ik\rightarrow j},\\
		\mathcal{T}_{k\rightarrow j} = \mathcal{U}_{k\rightarrow j} + \mathcal{R}_{ik\rightarrow j} .
\end{align}
\end{subequations}

{In the PID defined above, the terms $\mathcal{U}_{i\rightarrow j}$ and $\mathcal{U}_{k\rightarrow j}$ quantify the parts of the information transferred to the target process $Y_j$, which are unique to the source processes $Y_i$ and $Y_k$, respectively, thus reflecting contributions to the predictability of the target that can be obtained from one of the sources alone, but not from the other source alone. Each of these unique contributions sums up with the redundant transfer $\mathcal{R}_{ik\rightarrow j}$ to yield the information transfer from one source to the target as is known from the classic Shannon information theory. Then, the term $\mathcal{S}_{ik\rightarrow j}$ refers to the synergy between the two sources while they transfer information to the target, intended as the information that is uniquely obtained taking the two sources $Y_i$ and $Y_k$ together, but not considering them alone.}
Compared to the IID defined in (\ref{eq:IID}), the PID (\ref{eq:PID}) has the advantage that it provides distinct non-negative measures of redundancy and synergy, thereby accounting for the possibility that redundancy and synergy may coexist as separate elements of information modification. 
Interestingly, the IID and PID defined in Equations (\ref{eq:IID}) and (\ref{eq:PID}) are related to each other in a way such that:
\begin{equation} \label{eq:NetSynergy}
		\mathcal{I}_{ik\rightarrow j} = \mathcal{S}_{ik\rightarrow j} - \mathcal{R}_{ik\rightarrow j} ,
\end{equation}
thus showing that the interaction TE is actually a measure of the `net' synergy manifested in the transfer of information from the two sources to the target.

An issue with the PID (\ref{eq:PID}) is that its constituent measures cannot be obtained through classic information theory simply subtracting conditional MI terms as done for the IID; an additional ingredient to the theory is needed to get a fourth defining equation to be added to (\ref{eq:PID}) for providing an unambiguous definition of $\mathcal{U}_{i\rightarrow j}$, $\mathcal{U}_{k\rightarrow j}$, $\mathcal{R}_{ik\rightarrow j}$ and $\mathcal{S}_{ik\rightarrow j}$.
While several PID definitions have been proposed arising from different conceptual definitions of redundancy and synergy \cite{harder2013bivariate,griffith2014intersection,bertschinger2014quantifying}, here, we make reference to the so-called minimum MI (MMI) PID \cite{barrett2015exploration}. According to the MMI PID, redundancy is defined as the minimum of the information provided by each individual source to the target. In terms of information transfer measured by the TE, this leads to the following definition of the redundant TE:
\begin{equation} \label{eq:MMIredundancy}
		\mathcal{R}_{ik\rightarrow j} = \min \{\mathcal{T}_{i\rightarrow j}, \mathcal{T}_{k\rightarrow j} \} .
\end{equation}

This choice satisfies the desirable property that the redundant TE is independent of the correlation between the source processes. Moreover, it has been shown that, if the observed processes have a joint Gaussian distribution, all previously-proposed PID formulations reduce to the MMI PID \cite{barrett2015exploration}.

\section{Multiscale Information Transfer Decomposition} \label{sec3}
\vspace{-6pt}

\subsection{Multiscale Representation of Multivariate Gaussian Processes} \label{sec3.1}

In the linear signal processing framework, the $M$-dimensional vector stochastic process \mbox{$\bm{Y}_n=[Y_{1,n}\cdots Y_{M,n}]^T$}
is classically described using a vector autoregressive (VAR) \mbox{model of order $p$}:
\begin{equation} \label{eq:VAR}
\bm{Y}_n = \sum_{k=1}^{p}{\mathbf{A}_k \bm{Y}_{n-k} + \bm{U}_n}
\end{equation}
where $A_k$ are $M\times M$ matrices of coefficients and
$\bm{U}_n=[U_{1,n}\cdots U_{M,n}]^T$ is a vector of $M$ zero mean Gaussian processes with
covariance matrix {\boldmath$\Sigma$} $\equiv$ $\mathbb{E}[\bm{U}_n\bm{U}_n^T]$ ($\mathbb{E}$ is the expectation operator).
To study the observed process $\bm{Y}$ at the temporal scale identified by the scale factor $\tau$, we apply the following transformation to each constituent process $Y_m, m=1,\ldots,M$:
\begin{equation} \label{eq:MSY}
	\bar{Y}_{m,n} = \sum_{l=0}^{q}{b_{l} Y_{m,n\tau-l}}.
\end{equation}

This rescaling operation corresponds to transforming the original process $\bm{Y}$ through a two-step procedure that consists of the following {filtering} and {downsampling} steps, yielding respectively the {processes $\tilde{\bm{Y}}$ and $\bar{\bm{Y}}$}: 

\begin{subequations} \label{eq:AvgDws}
\begin{align}
		\tilde{\bm{Y}}_n &= \sum_{l=0}^{q}{b_l \bm{Y}_{n-l}} , \label{eq:Avg} \\
		\bar{\bm{Y}}_n &= \tilde{\bm{Y}}_{n\tau} , n=1,\ldots,N/\tau \label{eq:Dws}
\end{align}
\end{subequations}

The change of scale in (\ref{eq:MSY}) generalizes the averaging procedure originally proposed in \cite{costa2002multiscale}, which sets $q=\tau-1$ and $b_l=1/\tau$ and, thus, realizes the step of filtering through the simple procedure of~averaging $\tau$ subsequent samples. To improve the elimination of the fast temporal scales, in this study, we follow the idea of \cite{Valencia20092202}, in which a more appropriate low pass filter than averaging is employed. Here, we identify the $b_l$ as the coefficients of a linear finite impulse response (FIR) low pass filter of~order $q$; the FIR filter is designed using the classic window method with the Hamming window \cite{oppenheimdigital}, setting the cutoff frequency at $f_{\tau}=1/2\tau$ in order to avoid aliasing in the subsequent downsampling step.
Substituting (\ref{eq:VAR}) in (\ref{eq:Avg}), the filtering step leads to the process representation:
\begin{equation} \label{eq:VARMAAvg}
\tilde{\bm{Y}}_n = \sum_{k=1}^{p}{\mathbf{A}_k \tilde{\bm{Y}}_{n-k}} + \sum_{l=0}^{q}{\mathbf{B}_l \bm{U}_{n-l}}
\end{equation}
where $\mathbf{B}_l= b_l \mathbf{I}_M$ ($\mathbf{I}_M$ is the $M\times M$ identity matrix). Hence, the change of scale introduces a moving average (MA) component of order $q$ in the original VAR$(p)$ process, transforming it into a VARMA$(p,q)$ process. As we will show in the next section, the downsampling step (\ref{eq:Dws}) keeps the VARMA representation, altering the model parameters.

\subsection{State Space Processes}
\vspace{-6pt}

\subsubsection{Formulation of State Space Models}
{State space models are models that make use of {state variables} to describe a system by a set of~first-order difference equations, rather than by one or more high-order difference equations \cite{hannan2012statistical, aoki2013state}}.
The general linear state space (SS) model describing an observed vector process $\bm{Y}$ {has the form}: 

\begin{subequations} \label{eq:eqSS}
\begin{align}
		\bm{X}_{n+1} &= \mathbf{A} \bm{X}_{n} + \bm{W}_{n} \label{eq:eqSSstate} \\
		\bm{Y}_n &= \mathbf{C} \bm{X}_{n} + \bm{V}_{n} \label{eq:eqSSobs}
\end{align}
\end{subequations}
where the state Equation (\ref{eq:eqSSstate}) describes the update of the $L$-dimensional state (unobserved) process through the $L \times L$ matrix $\mathbf{A}$, and the observation Equation (\ref{eq:eqSSobs})
describes the instantaneous mapping from the state to the observed process through the $M \times L$ matrix $\mathbf{C}$. $\bm{W}_n$ and $\bm{V}_n$ are zero-mean white noise processes with covariances
$\mathbf{Q}$ $\equiv$ $\mathbb{E}[\bm{W}_n\bm{W}_n^T]$ and $\mathbf{R}$ $\equiv$ $\mathbb{E}[\bm{V}_n\bm{V}_n^T]$ and cross-covariance {\boldmath$S$} $\equiv$ $\mathbb{E}[\bm{W}_n\bm{V}_n^T]$. Thus, the parameters of the SS model (\ref{eq:eqSS}) are ($\mathbf{A},\mathbf{C},\mathbf{Q},\mathbf{R},\mathbf{S}$).

Another possible SS representation is that evidencing the {innovations} 
$\bm{E}_n=\bm{Y}_n-\mathbb{E}[\bm{Y}_n|\bm{Y}_n^-]$, i.e., the residuals of the linear regression of $\bm{Y}_n$ on its infinite past
$\bm{Y}_n^- = [\bm{Y}_{n-1}^T \bm{Y}_{n-2}^T \cdots]^T$ \cite{aoki2013state}. This new SS representation, usually referred to as the ``innovations form'' SS model (ISS), is characterized by the state process
$\bm{Z}_n=\mathbb{E}[\bm{X}_n|\bm{Y}_n^-]$ and by the $L \times M$ Kalman gain matrix $\mathbf{K}$:
\begin{subequations} \label{eq:eqISS}
\begin{align}
		\bm{Z}_{n+1} &= \mathbf{A} \bm{Z}_{n} + \mathbf{K} \bm{E}_{n} \label{eq:eqISSstate} \\
		\bm{Y}_n &= \mathbf{C} \bm{Z}_{n} + \bm{E}_{n} \label{eq:eqISSobs}
\end{align}
\end{subequations}

The parameters of the ISS model (\ref{eq:eqISS}) are ($\mathbf{A},\mathbf{C},\mathbf{K},\mathbf{V}$), 
where $\mathbf{V}$ is the covariance of the innovations, $\mathbf{V}$ $\equiv$ $\mathbb{E}[\bm{E}_n\bm{E}_n^T]$.
Note that the ISS (\ref{eq:eqISS}) is a special case of (\ref{eq:eqSS}) in which $\bm{W}_n=\mathbf{K} \bm{E}_n$ and $\bm{V}_n=\bm{E}_n$, so that $\mathbf{Q}=\mathbf{K}\mathbf{V}\mathbf{K}^T$, $\mathbf{R}=\mathbf{V}$ and $\mathbf{S}=\mathbf{K}\mathbf{V}$.

Given an SS model in the form (\ref{eq:eqSS}), the corresponding ISS model (\ref{eq:eqISS}) can be identified
by solving a so-called discrete algebraic Riccati equation ({DARE}) formulated in terms of the state error variance matrix $\mathbf{P}$ \cite{solo2016state}:
\begin{equation} \label{eq:DARE}
\begin{aligned}
	\mathbf{P} &= \mathbf{A}\mathbf{P}\mathbf{A}^T + \mathbf{Q} - (\mathbf{A}\mathbf{P}\mathbf{C}^T+\mathbf{S})
				(\mathbf{C}\mathbf{P}\mathbf{C}^T+\mathbf{R})^{-1} (\mathbf{C}\mathbf{P}\mathbf{A}^T+\mathbf{S}^T)
\end{aligned}
\end{equation}

Under some assumptions \cite{solo2016state}, the {DARE} (\ref{eq:DARE}) has {a unique} stabilizing solution, from which the Kalman gain and innovation covariance can be computed as:
\begin{equation} \label{eq:SStoISS}
\begin{aligned}
						\mathbf{V} &= \mathbf{C}\mathbf{P}\mathbf{C}^T + \mathbf{R} \\
						\mathbf{K} &= (\mathbf{A}\mathbf{P}\mathbf{C}^T + \mathbf{S})\mathbf{V}^{-1} ,
\end{aligned}
\end{equation}
thus completing the transformation from the SS form to the ISS form.

\subsubsection{State Space Models of Filtered and Downsampled Linear Processes}

Exploiting the close relation between VARMA models and SS models, first we show how to convert the VARMA model
(\ref{eq:VARMAAvg}) into an ISS model in the form of (\ref{eq:eqISS}) that describes the 
filtered process $\tilde{\bm{Y}_n}$. To do this, we exploit Aoki's method \cite{Aoki1991} defining the state process $\tilde{\bm{Z}}_n=[\bm{Y}_{n-1}^T \cdots \bm{Y}_{n-p}^T \bm{U}_{n-1}^T \cdots \bm{U}_{n-q}^T]^T$ that, together with $\tilde{\bm{Y}_n}$, obeys the state Equation
(\ref{eq:eqISS}) with parameters ($\tilde{\mathbf{A}},\tilde{\mathbf{C}},
\tilde{\mathbf{K}},\tilde{\mathbf{V}}$), where:
\[\tilde{\mathbf{A}}
=
\begin{bmatrix}
  \mathbf{A}_1&\cdots&\mathbf{A}_{p-1}&\mathbf{A}_p & \mathbf{B}_1&\cdots&\mathbf{B}_{q-1}&\mathbf{B}_q \\
		\mathbf{I}_M&\cdots&\mathbf{0}_M  &\mathbf{0}_M & \mathbf{0}_M&\cdots&\mathbf{0}_M  &\mathbf{0}_M \\
		\vdots   &   &\vdots     &\vdots   & \vdots   &   &\vdots     &\vdots		  \\
		\mathbf{0}_M&\cdots&\mathbf{I}_M  &\mathbf{0}_M & \mathbf{0}_M&\cdots&\mathbf{0}_M  &\mathbf{0}_M \\
		\mathbf{0}_M&\cdots&\mathbf{0}_M  &\mathbf{0}_M & \mathbf{0}_M&\cdots&\mathbf{0}_M  &\mathbf{0}_M \\
		\mathbf{0}_M&\cdots&\mathbf{0}_M  &\mathbf{0}_M & \mathbf{I}_M&\cdots&\mathbf{0}_M  &\mathbf{0}_M \\
		\vdots   &   &\vdots     &\vdots   & \vdots   &   &\vdots     &\vdots		  \\
		\mathbf{0}_M&\cdots&\mathbf{0}_M  &\mathbf{0}_M & \mathbf{0}_M&\cdots&\mathbf{I}_M  &\mathbf{0}_M
\end{bmatrix}
\]
\[\tilde{\mathbf{C}}
=
\begin{bmatrix}
	\mathbf{A}_1&\cdots&\mathbf{A}_p & \mathbf{B}_1&\cdots&\mathbf{B}_{q}
\end{bmatrix}
\]
\[\tilde{\mathbf{K}}
=
\begin{bmatrix}
	\mathbf{I}_M & \mathbf{0}_{M\times M(p-1)} &\mathbf{B}_0^{-T} & \mathbf{0}_{M\times M(q-1)}
\end{bmatrix}^T
\]
and $\tilde{\mathbf{V}} = \mathbf{B}_0$ {\boldmath$\Sigma$} $\mathbf{B}_0^T$, where $\tilde{\mathbf{V}}$
is the covariance of the innovations $\tilde{\bm{E}}_n=\mathbf{B}_0 \bm{U}_n$.

Now, we turn to show how the downsampled process $\bar{\bm{Y}}_n$ can be represented through an ISS model directly from the ISS formulation of the filtered process $\tilde{\bm{Y}}_n$. To this end, we exploit recent theoretical findings providing the state space form of downsampled signals (Theorem III in \cite{solo2016state}). Accordingly, the SS representation of the process downsampled at scale $\tau$, $\bar{\bm{Y}}_n=\tilde{\bm{Y}}_{n\tau}$ has parameters ($\bar{\mathbf{A}},\bar{\mathbf{C}},\bar{\mathbf{Q}},\bar{\mathbf{R}},\bar{\mathbf{S}}$), where $\bar{\mathbf{A}}=\tilde{\mathbf{A}}^\tau$, $\bar{\mathbf{C}}=\tilde{\mathbf{C}}$, 
$\bar{\mathbf{Q}}=\mathbf{Q}_\tau$, $\bar{\mathbf{R}}=\tilde{\mathbf{V}}$ and $\bar{\mathbf{S}}=\mathbf{S}_\tau$, with $\mathbf{Q}_\tau$ and $\mathbf{S}_\tau$ given by:
\begin{equation} \label{eq:QSdown}
\begin{aligned}
	\mathbf{S}_\tau &= \tilde{\mathbf{A}}^{\tau-1} \tilde{\mathbf{K}}\tilde{\mathbf{V}} \\
	\mathbf{Q}_\tau &= \tilde{\mathbf{A}} \mathbf{Q}_{\tau-1} \tilde{\mathbf{A}}^T
	+ \tilde{\mathbf{K}}\tilde{\mathbf{V}}\tilde{\mathbf{K}}^T, \tau\geq 2 \\
	\mathbf{Q}_1 &= \tilde{\mathbf{K}} \tilde{\mathbf{V}} \tilde{\mathbf{K}}^T, \tau=1 .
\end{aligned}
\end{equation}

Therefore, the downsampled process has an ISS representation with state process $\bar{\bm{Z}}_n=\tilde{\bm{Z}}_{n\tau}$, innovation process $\bar{\bm{E}}_n=\tilde{\bm{E}}_{n\tau}$ and parameters
($\bar{\mathbf{A}},\bar{\mathbf{C}},\bar{\mathbf{K}},\bar{\mathbf{V}}$), where $\bar{\mathbf{K}}$ and $\bar{\mathbf{V}}$ are obtained solving the {DARE} (\ref{eq:DARE}) and (\ref{eq:SStoISS}) for the SS model with parameters ($\bar{\mathbf{A}},\bar{\mathbf{C}},\bar{\mathbf{Q}},\bar{\mathbf{R}},\bar{\mathbf{S}}$).

\begin{figure}[H]
\centering
\includegraphics[width=14 cm]{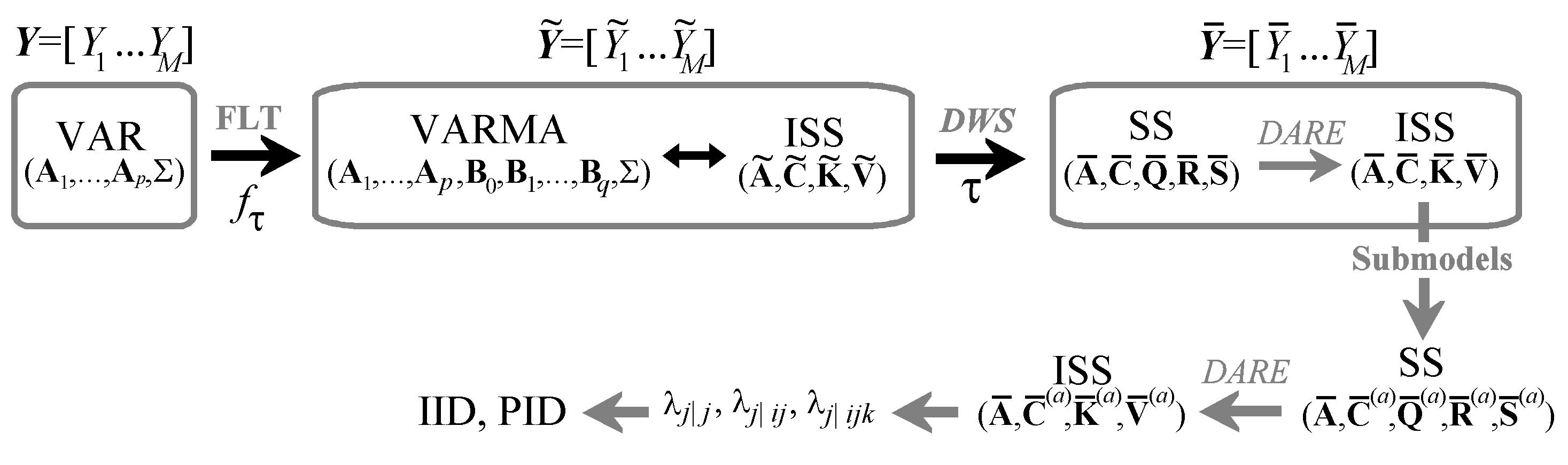}
\caption{Schematic representation of a linear VAR process and of its multiscale representation obtained through filtering (FLT) and downsampling (DWS) steps. The downsampled process has an innovations form state space model (ISS) representation from which submodels can be formed {to compute} the partial variances needed for the computation of information measures appearing in the IID and PID decompositions. This~makes it possible to perform multiscale information decomposition analytically from the original VAR parameters and from the scale factor.}
\label{fig2}
\end{figure}

To sum up, the relations and parametric representations of the original process $\bm{Y}$, the filtered process $\tilde{\bm{Y}}$ and the downsampled process $\bar{\bm{Y}}$ are depicted in Figure \ref{fig2}a. 
The step of low pass filtering (FLT) applied to a VAR($p$) process yields a VARMA($p,q$) process (where $q$ is the filter order, and the cutoff frequency is $f_\tau=1/2\tau$); this process is equivalent to an ISS process \cite{Aoki1991}. The subsequent downsampling (DWS) yields a different SS process, which in turn can be converted to the ISS form solving the {DARE}. Thus, both the filtered process $\tilde{\bm{Y}}_n$ and the downsampled process $\bar{\bm{Y}}_n$ can be represented as ISS processes with parameters 
($\tilde{\mathbf{A}},\tilde{\mathbf{C}},\tilde{\mathbf{K}},\tilde{\mathbf{V}}$) 
and ($\bar{\mathbf{A}},\bar{\mathbf{C}},\bar{\mathbf{K}},\bar{\mathbf{V}}$)
which can be derived analytically from the knowledge of the parameters of the original process ($\mathbf{A}_1,\ldots, \mathbf{A}_p, \Sigma$) and of the filter ($q, f_\tau$).
In the next section, we show how to compute analytically any measure appearing in the information decomposition of a jointly Gaussian multivariate stochastic process starting from its associated ISS model parameters, thus opening the way to the analytical computation of these measures for multiscale (filtered and downsampled) processes. 

\subsection{Multiscale IID and PID}

{After introducing the general theory of information decomposition and deriving the multiscale representation of the parameters of a linear VAR model,} in this section, we provide expressions for the terms of the IID and PID decompositions of the information transfer valid for multivariate jointly Gaussian processes.
The derivations are based on the knowledge that the linear parametric representation of Gaussian processes given in (\ref{eq:VAR}) captures all of the entropy differences that define the various information measures \cite{barrett2010multivariate} and that these entropy differences are related to the {partial variances} of the present of the target given its past and the past of one or more sources, intended as variances of the prediction errors resulting from linear regression \cite{faes2015information, faes2017information}.
Specifically, let us denote as $E_{j|j,n}=Y_{j,n}-\mathbb{E}[Y_{j,n}|Y_{j,n}^-]$,
$E_{j|ij,n}=Y_{j,n}-\mathbb{E}[Y_{j,n}|Y_{i,n}^-,Y_{j,n}^-]$ the prediction error of a linear regression of $Y_{j,n}$ performed respectively on $Y_{j,n}^-$ and $(Y_{j,n}^-,Y_{i,n}^-)$ and as $\lambda_{j|j}=\mathbb{E}[E_{j|j,n}^2]$, $\lambda_{j|ij}=\mathbb{E}[E_{j|ij,n}^2]$, the~corresponding prediction error variances. Then, the TE from $Y_i$ to $Y_j$ can be expressed as:
\begin{equation} \label{eq:TEgauss}
		T_{i\rightarrow j} = \frac{1}{2} \ln \frac {\lambda_{j|j}}{\lambda_{j|ij}} .
\end{equation}

In a similar way, the joint TE from $(Y_i,Y_k)$ to $Y_j$ can be defined as:
\begin{equation} \label{eq:JTEgauss}
		T_{ik\rightarrow j} = \frac{1}{2} \ln \frac {\lambda_{j|j}}{\lambda_{j|ijk}} ,
\end{equation}
where $\lambda_{j|ijk}=\mathbb{E}[E_{j|ijk,n}^2]$ is the variance of the prediction error of a linear regression of $Y_{j,n}$ on $(Y_{j,n}^-,Y_{i,n}^-,Y_{k,n}^-)$, $E_{j|ijk,n}=Y_{j,n}-\mathbb{E}[Y_{j,n}|Y_{i,n}^-,Y_{j,n}^-,Y_{k,n}^-]$. Based on these derivations, one can easily complete the IID decomposition of TE by computing $T_{k\rightarrow j}$ as in (\ref{eq:TEgauss}) and deriving the interaction TE from (\ref{eq:IID})
and the PID decomposition, as well by deriving the redundant TE from (\ref{eq:MMIredundancy}), the synergistic TE from (\ref{eq:NetSynergy}) and the unique TEs from (\ref{eq:PID}).

Next, we show how to compute any partial variance from the parameters of an ISS model in the form of (\ref{eq:eqISS}) \cite{barnett2015granger, solo2016state}. The partial variance $\lambda_{j|a}$, where the subscript $a$ denotes any combination of~indexes $\in \{1,\ldots,M\}$, can be derived from the ISS representation of the innovations of a {submodel} obtained removing the variables not indexed by $a$ from the observation equation. Specifically, we need to consider the submodel with state Equation (\ref{eq:eqISSobs}) and observation equation:
\begin{equation} \label{eq:SSreduced}
		Y_{n}^{(a)}=\mathbf{C}^{(a)} Z_n + E_{n}^{(a)} , 
\end{equation}
where the superscript $^{(a)}$ denotes the selection of the rows with indices $a$ of a vector or a matrix. It is important to note that the submodels (\ref{eq:eqISSstate}) and \eqref{eq:SSreduced} are {not} in innovations form, but are rather an SS model with parameters ($\mathbf{A},\mathbf{C}^{(a)},\mathbf{K}\mathbf{V}\mathbf{K}^T,
\mathbf{V}(a,a),\mathbf{K}\mathbf{V}(:,a)$). This SS model can be converted to an ISS model with innovation covariance $\mathbf{V}^{(a)}$ solving the {DARE} (\ref{eq:DARE}) and (\ref{eq:SStoISS}), so that the partial variance $\lambda_{j|a}$ is derived as the diagonal element of $\mathbf{V}^{(a)}$ corresponding to the position of the target $Y_j$. Thus, with this procedure, it is possible to compute the partial variances needed for the computation of the information measures starting from a set of ISS model parameters; since any VAR process can be represented at scale $\tau$ as an ISS process, the procedure allows computing the IID and PID information decompositions for the rescaled multivariate process (see Figure \ref{fig2}).

{It is worth remarking that, while the general formulation of IID and PID decompositions introduced in Section \ref{sec2} holds for arbitrary processes, the multiscale extension detailed in Section~\ref{sec3} is exact only if the processes have a joint Gaussian distribution. In such a case, the linear VAR representation captures exhaustively the joint variability of the processes, and any nonlinear extension has no additional utility (a formal proof of the fact that a stationary Gaussian VAR process must be linear can be found in \cite{barrett2010multivariate}). If, on the contrary, non-Gaussian processes are under scrutiny, the linear representation provided in Section \ref{sec3.1} can still be adopted, but may miss important properties in the dynamics and thus provide only a partial description. Moreover, since the close correspondence between conditional entropies and partial variances reported in this subsection does not hold anymore for non-Gaussian processes, all of the obtained measures should be regarded as indexes of (linear) predictability rather than as information measures.}

\section{Simulation Experiment}
To study the multiscale patterns of information transfer in a controlled setting with known dynamical interactions between time series, we consider a simulation scheme similar to some already used for the assessment of theoretical values of information dynamics \cite{faes2015information, faes2017information}.
Specifically, we analyze the following VAR process of order $M$ = 4:
\begin{subequations} \label{eq:simuVAR}
\begin{align}
		Y_{1,n} &= 2\rho_1 cos 2\pi f_1 Y_{1,n-1} - \rho_1^2 Y_{1,n-2} + U_{1,n} ,\\
		Y_{2,n} &= 2\rho_2 cos 2\pi f_2 Y_{2,n-1} - \rho_2^2 Y_{2,n-2} + cY_{1,n-1} + U_{2,n} ,\\
		Y_{3,n} &= 2\rho_3 cos 2\pi f_3 Y_{3,n-1} - \rho_3^2 Y_{3,n-2} + cY_{1,n-1} + U_{3,n} ,\\
		Y_{4,n} &= bY_{2,n-1} + (1-b)Y_{3,n-1} + U_{4,n} ,
\end{align}
\end{subequations}
where $\bm{U}_n=[U_{1,n}\cdots U_{4,n}]^T$ is a vector of zero mean white Gaussian noises with unit variance and uncorrelated with each other ({\boldmath$\Sigma$}= {\boldmath$I$}). 
The parameter design in Equation (\ref{eq:simuVAR}) is chosen to allow autonomous oscillations in the processes $Y_i$, $i=1,\ldots, 3$, obtained placing complex-conjugate poles with modulus $\rho_i$ and frequency $f_i$ in the complex plane representation of the transfer function \mbox{of the vector} process, as~well as causal interactions between the processes at a fixed time lag of one sample and with strength modulated by the parameters $b$ and $c$ (see Figure \ref{sec3}).
In this study, we set the coefficients related to self-dependencies to values generating well-defined oscillations in all processes \mbox{($\rho_1=\rho_2=\rho_3=0.95$)} and letting $Y_1$ fluctuate at slower time scales than $Y_2$ and $Y_3$ ($f_1=0.1, f_2=f_3=0.025$).
We consider four configurations of the parameters, chosen to reproduce paradigmatic conditions of interaction between the processes:
\begin{enumerate}[label=(\alph*),leftmargin=2.3em,labelsep=4mm]
 \item isolation of $Y_1$ and $Y_2$ and unidirectional coupling $Y_3 \rightarrow Y_4$, obtained setting $b=c=0$;
 \item common driver effects $Y_2 \leftarrow Y_1 \rightarrow Y_3$ and unidirectional coupling $Y_3 \rightarrow Y_4$, obtained setting $b=0$ and $c=1$;
 \item isolation of $Y_1$ and unidirectional couplings $Y_2 \rightarrow Y_4$ and $Y_3 \rightarrow Y_4$, obtained setting $b=0.5$ and~$c=0$;
	\item common driver effects $Y_2 \leftarrow Y_1 \rightarrow Y_3$ and unidirectional couplings $Y_2 \rightarrow Y_4$ and $Y_3 \rightarrow Y_4$, obtained setting $b=0.5$ and $c=1$.
\end{enumerate}

\begin{figure}[H]
\centering
\includegraphics[width=15 cm]{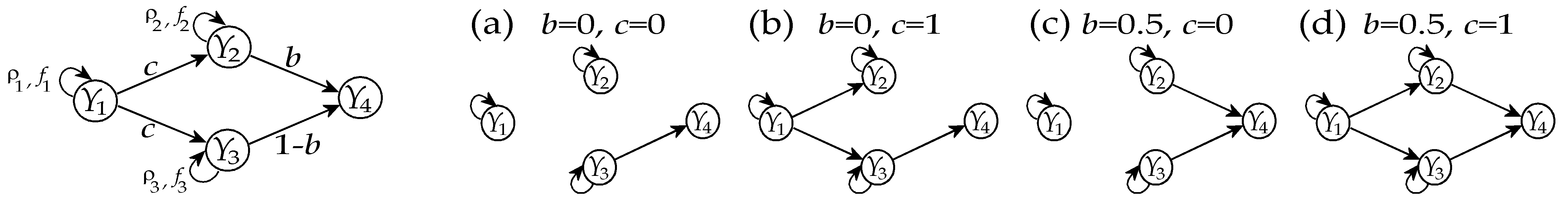}
\caption{Graphical representation of the four-variate VAR process of Equation (\ref{eq:simuVAR}) that we use to explore the multiscale decomposition of the information transferred to $Y_4$, selected as the target process, from~$Y_2$ and $Y_3$, selected as the source processes, in the presence of $Y_1$, acting as the exogenous process. To~favor such exploration, we set oscillations at different time scales for $Y_1$ ($f_1=0.1$) and for $Y_2$ and $Y_3$ ($f_2=f_3=0.025$), induce common driver effects from the exogenous process to the sources modulated by the parameter $c$ and allow for varying strengths of the causal interactions from the sources to the target as modulated by the parameter $b$. The four configurations explored in this study are depicted in (\textbf{a}--\textbf{d}).}
\label{fig3}
\end{figure}

\begin{figure}[H]
\centering
\includegraphics[width=14 cm]{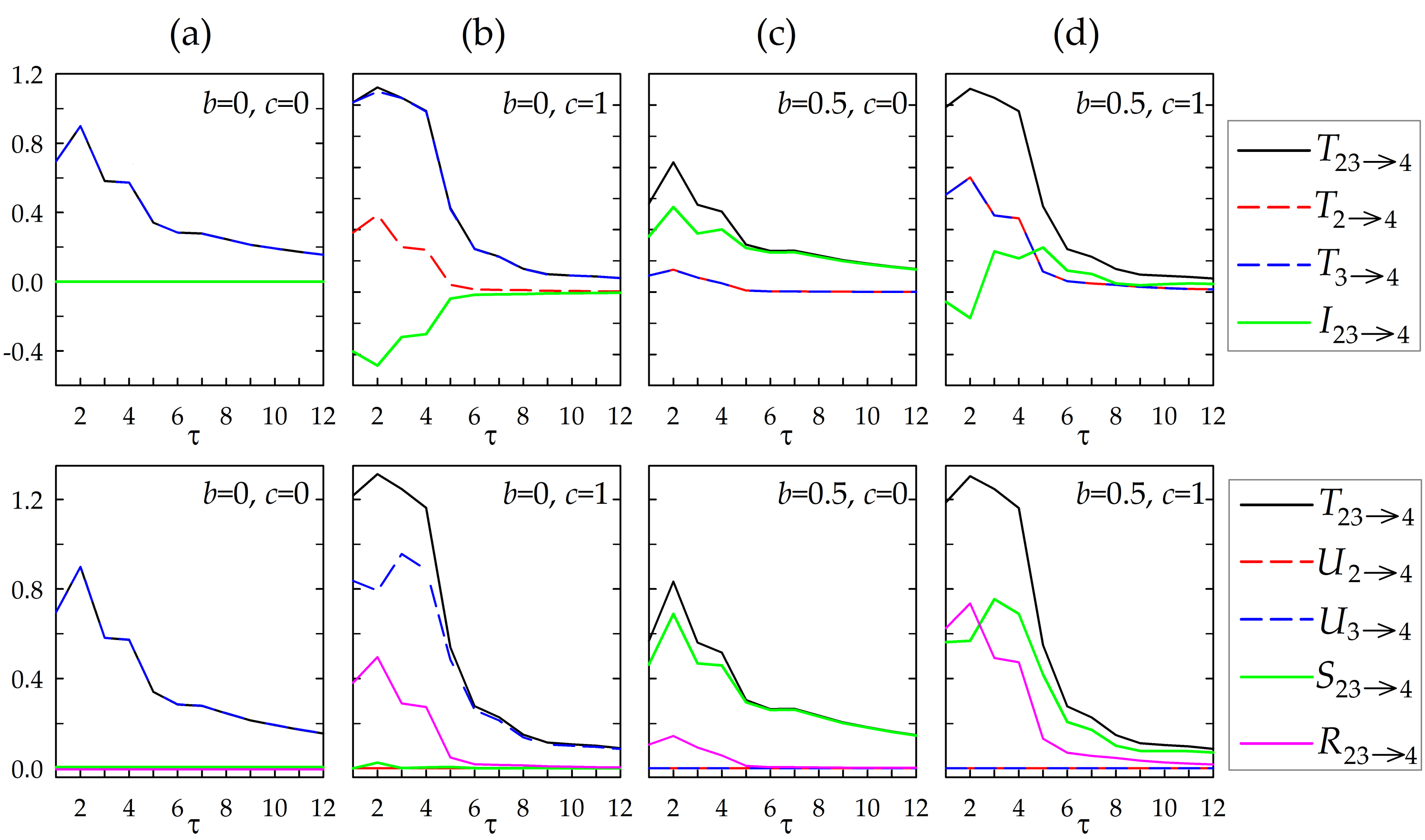}
\caption{Multiscale information decomposition for the simulated VAR process of Equation (20). Plots depict the exact values of the entropy measures forming the interaction information decomposition (IID, upper row) and the partial information decomposition (PID, lower row) of the information transferred from the source processes $Y_2$ and $Y_3$ to the target process $Y_4$ generated according to the scheme of Figure \ref{fig3} with four different configurations of the parameters. We find that linear processes may generate trivial information patterns with the absence of synergistic or redundant behaviors (\textbf{a}), patterns with the prevalence of redundant information transfer (\textbf{b}) or synergistic information transfer (\textbf{c}) that persist across multiple time scales, or even complex patterns with the alternating prevalence of~redundant transfer and synergistic transfer at different time scales (\textbf{d}).}
\label{fig4}
\end{figure}

With this simulation setting, we compute all measures appearing in the IID and PID decompositions of the information transfer, considering $Y_4$ as the target process and $Y_2$ and $Y_3$ as the source processes.
The theoretical values of these measures, computed as a function of the time scale using the IID and the PID, are reported in Figure \ref{fig4}.
In the simple case of unidirectional coupling $Y_3 \rightarrow Y_4$ ($b=c=0$, Figure \ref{fig4}a), the joint information transferred from $(Y_2,Y_3)$ to $Y_4$ is {exclusively due to} the source $Y_3$ without contributions from $Y_2$ and without interaction effects between the sources $(\mathcal{T}_{23\rightarrow 4}=\mathcal{T}_{3\rightarrow 4}=\mathcal{U}_{3\rightarrow 4}, \mathcal{T}_{2\rightarrow 4}=\mathcal{U}_{2\rightarrow 4}=0, \mathcal{I}_{23\rightarrow 4}=\mathcal{S}_{23\rightarrow 4}=\mathcal{R}_{23\rightarrow 4}=0)$.

When the causal interactions towards $Y_4$ are still due exclusively to $Y_3$, but the two sources $Y_2$ and $Y_3$ share information arriving from $Y_1$ ($b=0, c=1$; Figure \ref{fig4}b), the IID evidences that the joint information transfer coincides again with the transfer from $Y_3$ ($\mathcal{T}_{23\rightarrow 4}=\mathcal{T}_{3\rightarrow 4}$), but a non-trivial amount of information transferred from $Y_2$ to $Y_4$ emerges, which is fully redundant ($\mathcal{T}_{2\rightarrow 4}=-\mathcal{I}_{23\rightarrow 4}$). The PID highlights that the information from $Y_3$ to $Y_4$ is not all unique, but is in part transferred redundantly with $Y_2$, while the unique transfer from $Y_2$ and the synergistic transfer are negligible.

In the case of two isolated sources equally contributing to the target ($b=0.5, c=0$, Figure \ref{fig4}c), the~IID evidences the presence of net synergy and of identical amounts of information transferred to $Y_4$ from $Y_2$ or $Y_3$ ($\mathcal{I}_{23\rightarrow 4}>0, \mathcal{T}_{2\rightarrow 4}=\mathcal{T}_{3\rightarrow 4}$). The PID documents that there are no unique contributions, so~that the two amounts of information transfer from each source to the target coincide with the redundant transfer, and the remaining part of the joint transfer is synergistic ($\mathcal{U}_{2\rightarrow 4}=\mathcal{U}_{3\rightarrow 4}=0$, \mbox{$\mathcal{T}_{2\rightarrow 4}=\mathcal{T}_{3\rightarrow 4}=\mathcal{R}_{23\rightarrow 4}, \mathcal{S}_{23\rightarrow 4}=\mathcal{T}_{23\rightarrow 4}-\mathcal{R}_{23\rightarrow 4}$}).

Finally, when the two sources share common information and contribute equally to the target ($b=0.5, c=1$; Figure \ref{fig4}d), we find that they send the same amount of information as before, but~in this case, no unique information is sent by any of the sources ($\mathcal{T}_{2\rightarrow 4}=\mathcal{T}_{3\rightarrow 4}, \mathcal{U}_{2\rightarrow 4}=\mathcal{U}_{3\rightarrow 4}=0$). Moreover, the nature of the interaction between the sources is not trivial and is scale dependent: at low time scales, where the dynamics are likely dominated by the fast oscillations of $Y_1$, the IID reveals net redundancy, and the PID shows that the redundant transfer prevails over the synergistic ($\mathcal{I}_{23\rightarrow 4}<0, \mathcal{R}_{23\rightarrow 4}>\mathcal{S}_{23\rightarrow 4}$); at higher time scales, where fast dynamics are filtered out and the slow dynamics of $Y_2$ and $Y_3$ prevail, the IID reveals net synergy, and the PID shows that the synergistic transfer prevails over the redundant ($\mathcal{I}_{23\rightarrow 4}>0, \mathcal{S}_{23\rightarrow 4}>\mathcal{R}_{23\rightarrow 4}$).




\section{Application}
As a real data application, we analyze intracranial EEG recordings from a patient with drug-resistant epilepsy measured by an implanted array of $8\times 8$ cortical electrodes and two left hippocampal depth electrodes with six contacts each. The data are available in \cite{epilepsy}, and further details on the dataset are
given in \cite{Kramer2008}. Data were sampled at 400 Hz and correspond to 10-s segments recorded in the pre-ictal period, just before the seizure onset, and 10 s during the ictal stage \mbox{of the seizure}, \mbox{for a total} of eight seizures.
Defining and locating the seizure onset zone, i.e., the specific location in the brain where the synchronous activity of neighboring groups of cells becomes so strong so as to be able to spread its own activity to other distant regions, is an important issue in the study of epilepsy in humans.
Here, we focus on the information flow from the sub-cortical regions, probed by depth electrodes, to the brain cortex. In \cite{stramaglia2014synergy}, it has been suggested that Contacts 11 and 12, in the second depth electrode, are mostly influencing the cortical activity; accordingly, in this work, we consider Channels 11 and 12 as a pair of source variables for all of the cortical electrodes and decompose the information flowing from them using the multiscale IID and PID here proposed, both in the pre-ictal stage and in the ictal stage. An FIR filter with $q=12$ coefficients is used, and the order $p$ of the VAR model is fixed according to the Bayesian information criterion. 
{In the analyzed dataset, the model order assessed in the pre-ictal phase was $p=14.61 \pm 1.07$ (mean $\pm$ std. dev.
 across 64 electrodes and eight~seizures) and during the ictal phase decreased significantly to $p=11.09 \pm 3.95$.}

\begin{figure}[H]
\centering
\includegraphics[width=14 cm]{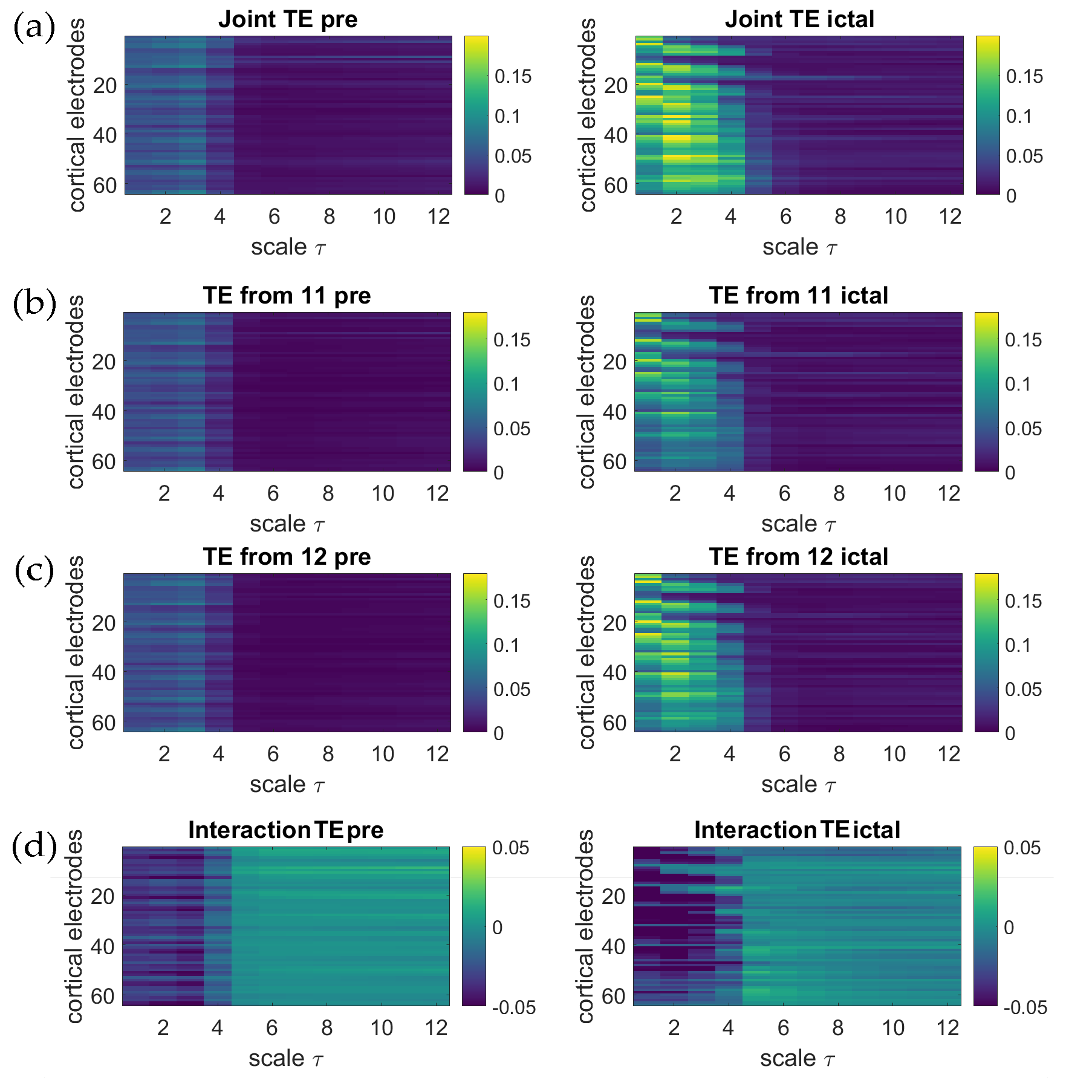}
\caption{Interaction information decomposition (IID) of the intracranial EEG information flow from subcortical to cortical regions in an epileptic patient. The joint transfer entropy from depth \mbox{Channels 11 and 12} to cortical electrodes (\textbf{a}); the transfer entropy from depth Channel 11 to cortical electrodes (\textbf{b}); the transfer entropy from depth Channel 12 to cortical electrodes (\textbf{c}) and the interaction transfer entropy from depth Channels 11 and 12 to cortical electrodes (\textbf{d}) are depicted as a function of the scale $\tau$, after averaging over the eight pre-ictal segments (left column) and over the eight ictal segments (right column). Compared with pre-ictal periods, during the seizure, the IID evidences marked increases of the joint and individual information transfer from depth to cortical electrodes and~low and almost unvaried levels of interaction transfer.}
\label{fig5}
\end{figure}

In Figure \ref{fig5}, we depict the terms of the IID applied from the two sources (Channels $\{11,12\}$) to any of the electrodes as a function of the scale $\tau$, averaged over the eight seizures. We observe a~relevant enhancement of the joint TE during the seizure, w.r.t. the pre-ictal period. This enhancement is determined by a marked increase of both the individual TEs from Channels 11 and 12 to all of the cortical electrodes; the patterns of the two TEs are similar to each other in both stages. The pattern of interaction information transfer displays prevalent redundant transfer for low values of $\tau$ and prevalent synergistic transfer for high $\tau$, but the values of the interaction TE have relatively low magnitude and are only slightly different in pre-ictal and ictal conditions. It is worth stressing that at scale $\tau$, the algorithm analyzes oscillations, in the time series, slower than $\frac{1}{2\tau f_s}$ s, where $f_s =400$ Hz.

\begin{figure}[H]
\centering
\includegraphics[width=14 cm]{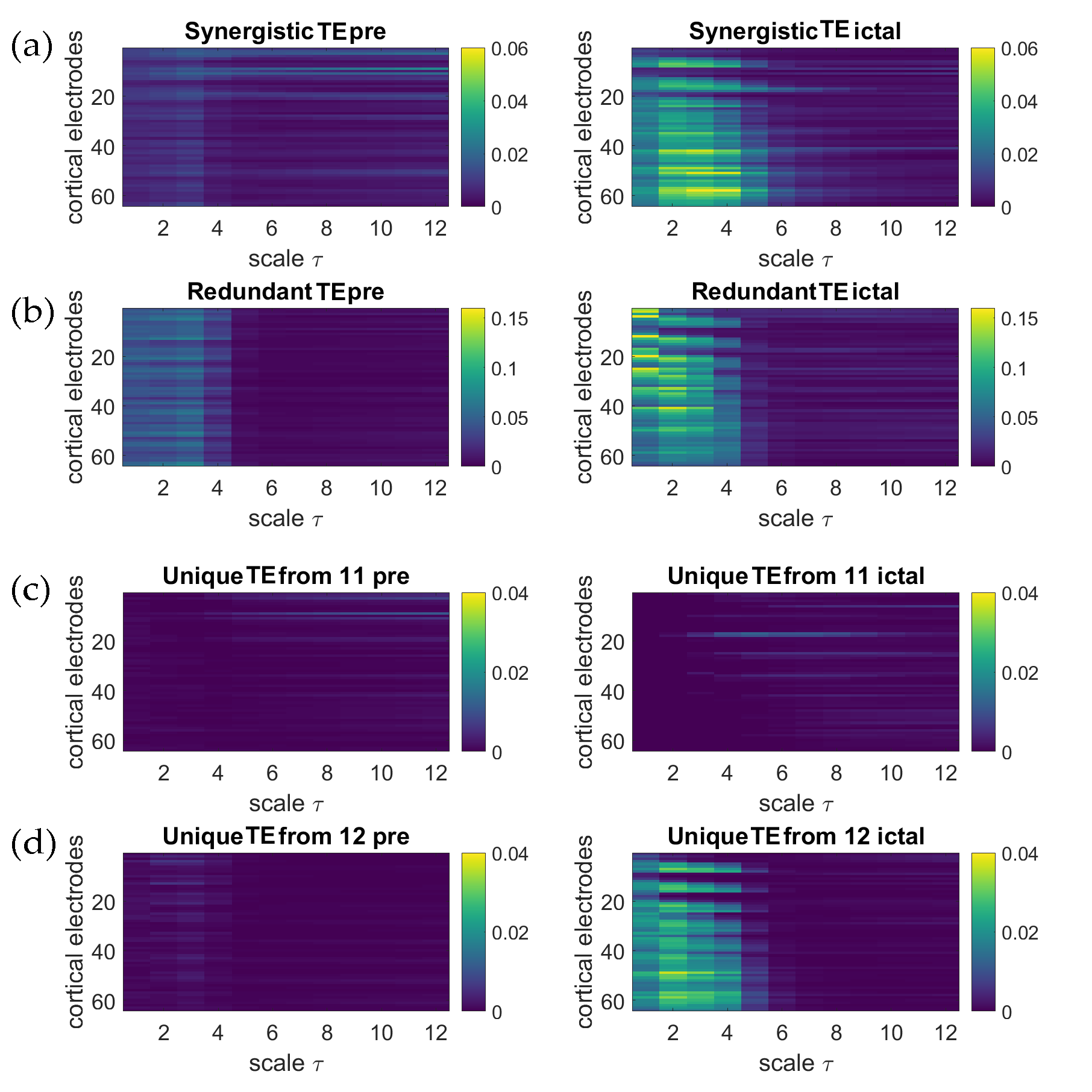}
\caption{Partial information decomposition (PID) of the intracranial EEG information flow from subcortical to cortical regions in an epileptic patient. The synergistic transfer entropy from depth Channels 11 and 12 to cortical electrodes (\textbf{a}); the redundant transfer entropy from depth \mbox{Channels 11 and 12} to cortical electrodes (\textbf{b}); the unique transfer entropy from depth Channel 11 to cortical electrodes (\textbf{c}) and the unique transfer entropy from depth {Channel 12} to cortical electrodes (\textbf{d})~are depicted as a function of the scale $\tau$, after averaging over the eight pre-ictal segments (left~column) and over the eight ictal segments (right column). Compared with pre-ictal periods, during the seizure, the PID evidences marked increases of the information transferred synergistically and redundantly {from depth} to cortical electrodes and of the information transferred uniquely from one of the two depth electrodes, but not from the other.}
\label{fig6}
\end{figure}

In Figure \ref{fig6}, we depict, on the other hand, the terms of the PID computed for the same data. This~decomposition shows that the increased joint TE across the seizure transition seen in Figure~\ref{fig5}a is in large part the result of an increase of both the synergistic and the redundant TE, which are markedly higher during the ictal stage compared with the pre-ictal. This explains why the interaction TE of~Figure~\ref{fig5}d, which is the difference between two quantities that both increase, is nearly {constant} moving from the pre-ictal to the ictal stage.
The quantity that, instead, clearly differentiates between \mbox{Channels~11 and 12} is the unique information transfer: indeed, only the unique TE from Channel~12 increases in the ictal stage, while the unique TE from Channel 13 remains at low levels.

\begin{figure}[H]
\centering
\includegraphics[width=15.3 cm]{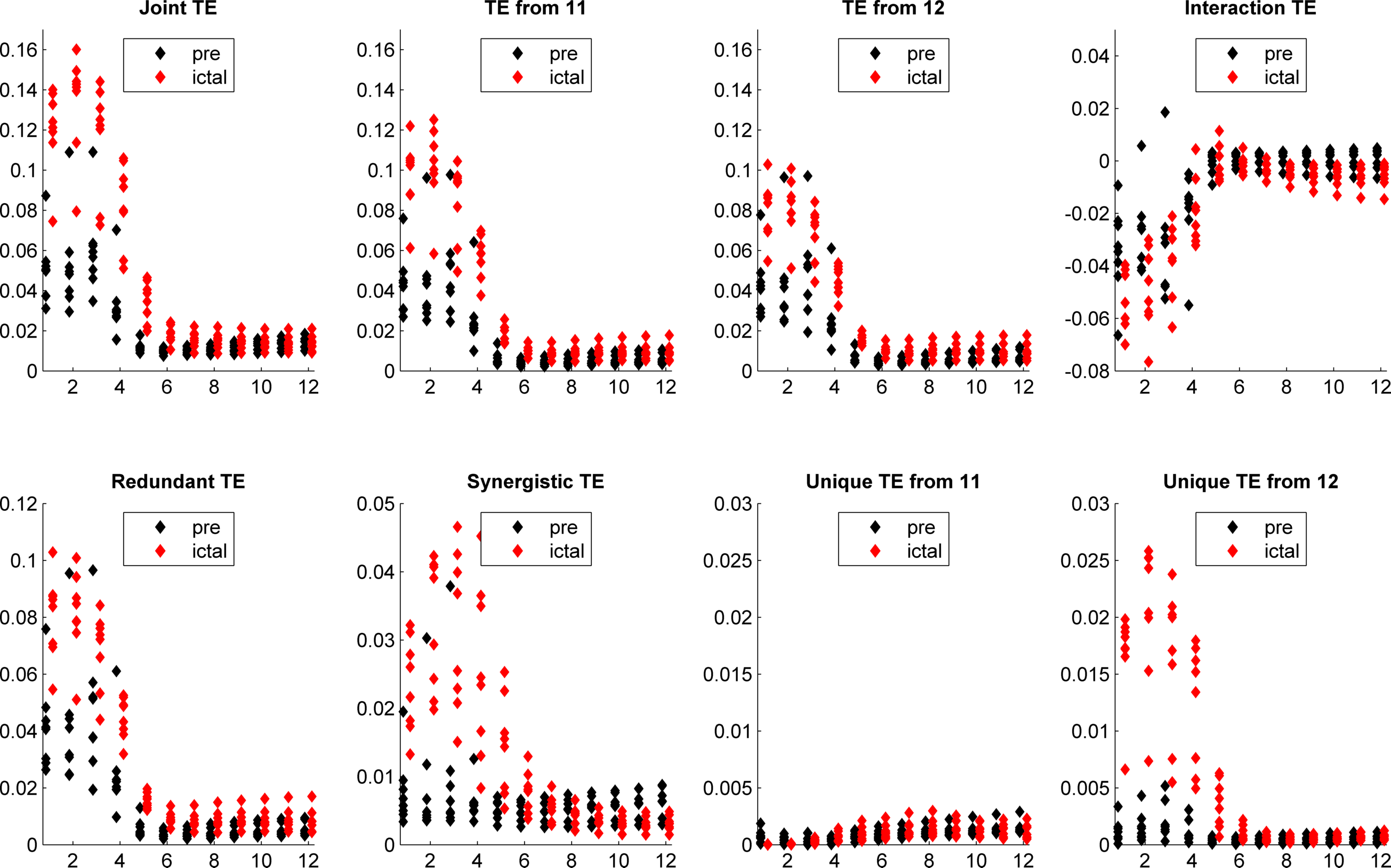}
\caption{Multiscale representation of the measures of interaction information decomposition (IID, {top}) and partial information decomposition (PID, {bottom}) computed as a function of the time scale for each of the eight seizures during the pre-ictal period (black) and the ictal period (red). Values of joint transfer entropy (TE), individual TE, interaction TE, redundant TE, synergistic TE and unique TE are obtained taking the depth Channels 11 and 12 as sources and averaging over all 64 target cortical electrodes.
Increases during seizure of the joint TE, individual TEs from both depth electrodes, redundant and synergistic TE and unique TE from the depth electrode 12 are evident at low time scales for almost all considered episodes.}
\label{fig7}
\end{figure}

{In order to investigate the variability across trials of the estimates of the various information measures, in Figure \ref{fig7}, we depict the terms of both IID and PID expressed for each ictal episode as average values over all 64 cortical electrodes. The analysis shows that the higher average values observed in Figures \ref{fig5} and \ref{fig6} at Scales 1--4 during the ictal state for the joint TE, the two individual TEs, the redundant and synergistic TEs and the unique TE from depth Channel 12 are the result of an~increase of the measures for almost all of the observed seizure episodes.}

{These findings are largely in agreement with the increasing awareness that epilepsy is a network phenomenon that involves aberrant functional connections across vast parts of the brain on virtually all spatial scales \cite{richardson2012large, dickten2016weighted}. Indeed, our} results document that the occurrence of seizures is associated with a relevant increase of the information flowing from the subcortical regions (associated with the depth electrode) to the cortex and that the character of this information flow is mostly redundant both in the pre-ictal and in the {ictal state}. Here, the need for a multiscale approach is testified by the fact that several quantities in the {ictal state} (e.g., the joint TE, the synergistic IT
 and the unique IT
 from Channel~12) attain their maximum at scale $\tau > 1$.

Moreover, {the approaches that we propose for information decomposition appear useful to improve the localization of epileptogenic areas in patients with drug-resistant epilepsy. Indeed,} our~analysis suggests that Contact 12 is the closest to the seizure onset zone, and it is driving the cortical oscillations during the ictal stage, as it sends unique information to the cortex. On the other hand, to~disentangle this effect, it has been necessary to include also Channel 11 in the analysis and to make the PID of the total information from the pair of depth channels to the cortex; indeed, the redundancy between Channels 11 and 12 confounds the informational pattern unless the PID is performed.

\section{Conclusions}
Understanding how multiple inputs may combine to create the output of a given target is a~fundamental challenge in many fields, in particular in neuroscience. Shannon's information theory is the most suitable frame to cope with this problem and thus to assess the informational character of multiplets of variables describing complex systems; IID indeed measures the balance between redundant and synergetic interaction within the classical multivariate entropy-based approach. Recently Shannon's information theory has been extended, in the PID, so as to provide specific measures for the information that several variables convey
individually (unique information), redundantly (shared information) or only jointly (synergistic information) about the output.

The contribution of the present work is the proposal of an analytical frame where both IID and PID can be exactly evaluated in a multiscale fashion, for multivariate Gaussian processes, on the basis of simple vector autoregressive identification. In doing this, our work opens the way for both the theoretical analysis and the practical implementation of information modification in processes that exhibit multiscale dynamical structures. 
The effectiveness of the proposed approach has been demonstrated both on simulated examples and on real publicly-available intracranial EEG data. Our~results provide a firm ground to the multiscale evaluation of PID, to be applied in all applications where causal influences coexist at multiple temporal scales. 

Future developments of this work include the refinement of the SS model structure to accommodate the description of long-range linear correlations \cite{sela2009computationally} or its expansion to the description of nonstationary processes \cite{kitagawa1987non} and the formalization of exact cross-scale computation of information decomposition within and between multivariate processes \cite{paluvs2014cross}.
A major challenge in the field remains the generalization of this type of analysis to non-Gaussian processes, for which exact analytical solutions or computationally-reliable estimation approaches are still lacking.
{This constitutes a~main direction for further research, because real-world processes display very often non-Gaussian distributions, which would make an extension to nonlinear models or model-free approaches beneficial. The questions that are still open in this respect include the evaluation of proper theoretical definitions of synergy or redundancy for nonlinear processes \cite{williams2010nonnegative,harder2013bivariate,griffith2014intersection,e19020085,bertschinger2014quantifying}, the development of reliable entropy estimators for multivariate variables with different dimensions \cite{papana2011reducing,wibral2014directed,Faes2015estimatingthe} and the assessment of the extent to which non-linear model-free methods really outperform the linear model-based approach adopted here and in previous investigations \cite{porta2017arenonlinear}.}

\end{document}